\pdfoutput=1
\documentclass[conference]{IEEEtran}

\makeatletter
\def\endthebibliography{%
  \def\@noitemerr{\@latex@warning{Empty `thebibliography' environment}}%
  \endlist
}
\makeatother
\usepackage{pgfplots}
\usepgfplotslibrary{groupplots}
\pgfplotsset{compat=1.18}
\usepackage{pgfplotstable}
\usepackage{graphicx}
\usepackage{subfig}
\usepackage{multirow}
\usepackage{lipsum}
\usepackage{xspace}
\usetikzlibrary{patterns}
\usepackage{amsmath}
\usepackage{amsfonts}
\usepackage{cite}
\usepackage{amssymb}

\newcommand*\circled[1]{\tikz[baseline=(char.base)]{
            \node[shape=circle,draw,inner sep=0.7pt] (char) {#1};}}
\def\BibTeX{{\rm B\kern-.05em{\sc i\kern-.025em b}\kern-.08em
    T\kern-.1667em\lower.7ex\hbox{E}\kern-.125emX}}
\usepackage{float}
\usepackage[hidelinks]{hyperref}
\usetikzlibrary{positioning}

\usepackage{algorithm}
\usepackage{algpseudocode}
\usepackage{pifont}% http://ctan.org/pkg/pifont
\usepackage{newunicodechar}
\usepackage{fancyhdr}
\newunicodechar{≤}{\ensuremath{\leq}}
\usepackage{pifont}% http://ctan.org/pkg/pifont
\usepackage{fancyhdr}
\usepackage{amsmath}
\usepackage{amssymb}
\makeatletter
\let\old@ps@headings\ps@headings
\let\old@ps@IEEEtitlepagestyle\ps@IEEEtitlepagestyle
\def\confheader#1{%
    \def\ps@headings{%
        \old@ps@headings%
        \def\@oddhead{\strut\hfill#1\hfill\strut}%
        \def\@evenhead{\strut\hfill#1\hfill\strut}%
    }%
    \def\ps@IEEEtitlepagestyle{%
        \old@ps@IEEEtitlepagestyle%
        \def\@oddfoot{\strut\hfill\thepage\hfill\strut}
        \def\@evenfoot{\strut\hfill\thepage\hfill\strut}
        \def\@oddhead{\strut\hfill#1\hfill\strut}%
        \def\@evenhead{\strut\hfill#1\hfill\strut}%
    }%
    \ps@headings%
}

\makeatother

\newif\ifassyst

\assystfalse % <--- Uncomment for the MLArchSys workshop

\ifassyst
  \confheader{%
        \centering{Architecture and System Support for Transformer Models (ASSYST), ISCA, 2023}
    }
\else
  \confheader{%
        \centering{ML for Computer Architecture and Systems (MLArchSys), ISCA, 2023}
}
\fi

\title{Towards Efficient Multi-Agent Learning Systems}
\author{Kailash Gogineni, %~\IEEEmembership{}
        Peng Wei, %~\IEEEmembership{Member,~IEEE, }% <-this % stops a space
        Tian Lan and %~\IEEEmembership{Member,~IEEE, }
        Guru Venkataramani\\%~\IEEEmembership{Senior Member,~IEEE}
The George Washington University, Washington, DC, USA\\%\protect\\
E-mail: \{kailashg26, pwei, tlan, guruv\}@gwu.edu}

\begin{document}
\maketitle

\pagestyle{plain}

%%%%%% -- PAPER CONTENT STARTS-- %%%%%%%%

\begin{abstract}
Multi-Agent Reinforcement Learning (MARL) is an increasingly important research field that can model and control multiple large-scale autonomous systems. Despite its achievements, existing multi-agent learning methods typically involve expensive computations in terms of training time and power arising from large observation-action space and a huge number of training steps. Therefore, a key challenge is understanding and characterizing the computationally intensive functions in several popular classes of MARL algorithms during their training phases. Our preliminary experiments reveal new insights into the key modules of MARL algorithms that limit the adoption of MARL in real-world systems. We explore neighbor sampling strategy to improve cache locality and observe performance improvement ranging from $26.66\%$ ($3$ agents) to $27.39\%$ ($12$ agents) during the computationally intensive mini-batch sampling phase. Additionally, we demonstrate that improving the locality leads to an end-to-end training time reduction of $10.2\%$~(for $12$ agents) compared to existing multi-agent algorithms without significant degradation in the mean reward.
\end{abstract}


\section{Introduction}\label{sec:introduction}
Reinforcement Learning~(RL) has recently made exciting progress in many applications, including Atari games~\cite{mnih2013playing}, aviation systems~\cite{razzaghi2022survey}, and robotics~\cite{wang2022equivariant}. Specifically, RL frameworks fit in the context of problems that involve sequential-decision making where the agent needs to take actions in an environment to maximize the cumulative rewards. In RL, the quality of state-action pairs is evaluated using a reward function, and the transition to a new state depends on the current state and action~\cite{sutton2018reinforcement}. The function that determines the action from the state is known as a policy. The function representing the reward estimates is known as the value function.

Multi-agent systems~\cite{sutton2018reinforcement} have shown excellent performance among various multi-player games~\cite{zhang2021multi} where there is significant sharing of observations between the agents during training, and the joint actions among these agents could affect the environment dynamically. In MARL, several agents simultaneously explore a common environment and perform competitive~\textcolor{black}{(e.g., Predator-prey) and cooperative~(e.g., Cooperative navigation) tasks~\cite{lowe2017multi}}. All the observations are shared in the cooperative setting, and the training is performed centrally. In contrast, each agent aims to outperform its enemies in a competitive setting. As a result, MARL training involves several {\it computationally-challenging} tasks that deal with dynamically changing environments.

\begin{figure}
    [!ht]\centering
    \includegraphics[scale=0.44]{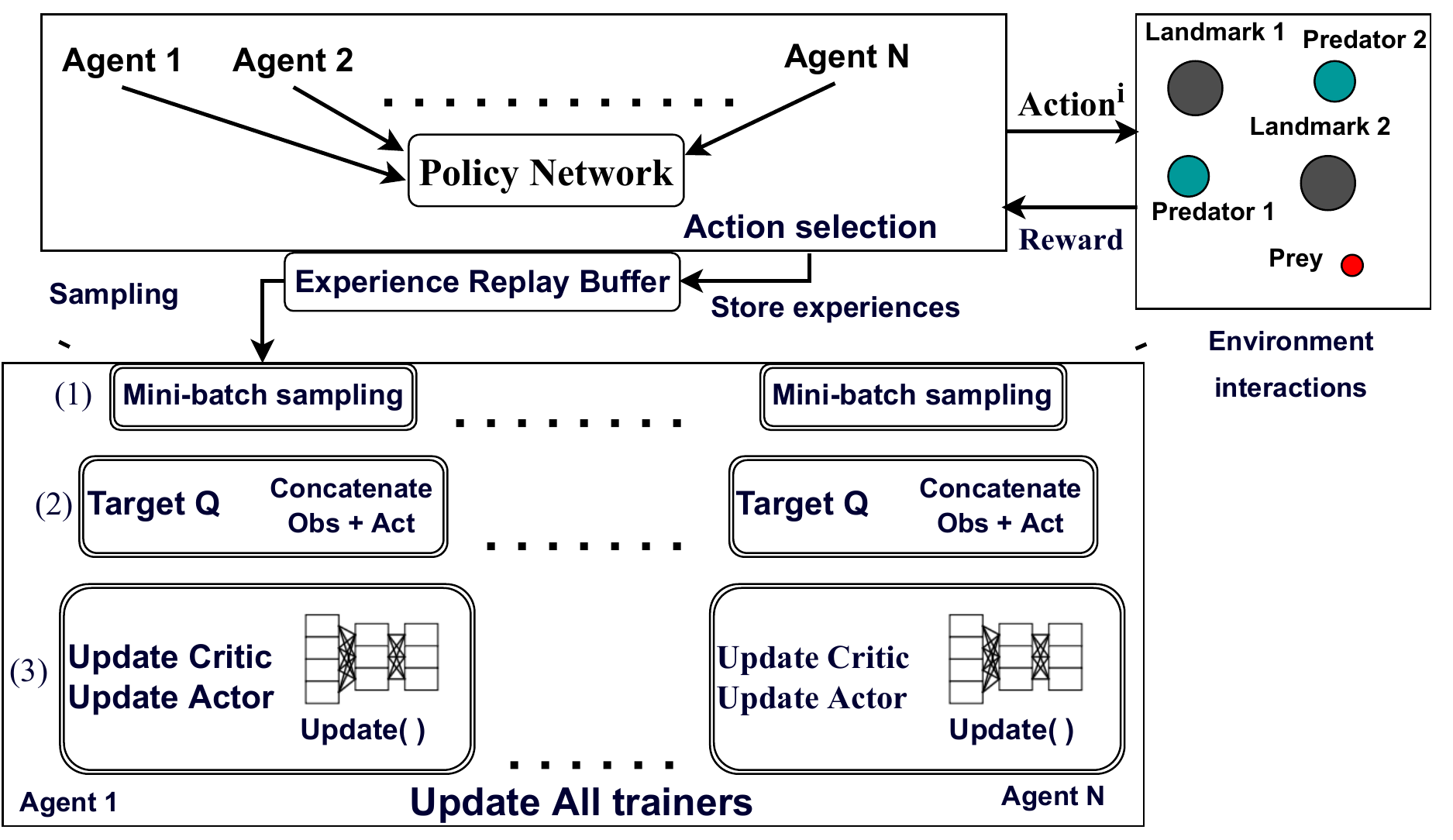}
    \caption{\textcolor{black}{Overview of our multi-agent decentralized actor, centralized critic approach~(Competitive environment).}}
  \label{figure1}
\end{figure}

In this paper, we performed a workload characterization study to understand the {\it performance-limiting functions} on well-known model-free MARL frameworks~\cite{lowe2017multi, haarnoja2018soft} implemented using actor-critic methods with state spaces that are usually very large. We analyze different MARL training phases where the actor and critic networks are responsible for policy and value functions, respectively. The critic tries to learn a value function given the policy from the actor, while the actor can estimate the policy gradient based on the approximate value function that the critic provides. 

As shown in Figure~\ref{figure1}, each agent in the environment has its own actor network which outputs the action of an agent given its observation~(\textit{Action selection}). During the \textit{mini-batch sampling} phase, each agent $i$ collects the historical transition data of all other agents stored within the \textit{Experience Replay Buffer}. The sampling approach enables the algorithm to reuse the transition data for updating the current policy. Each agent has a centralized critic which outputs the Q-value using the joint observation-action space of all other agents. During \textit{Update all trainers} phase, both the actor and critic networks are updated after the \textit{target Q calculation} and \textit{sampling phase}. At last, we explore the neighbor sampling strategy for better cache locality and that leads to a performance improvement ranging from $26.66\%$ ($3$ agents) to $27.39\%$ ($12$ agents)~(Figure~\ref{figure6}). 
 
\begin{figure*}
    [ht]\centering
    \includegraphics[width=14cm, height=7cm]{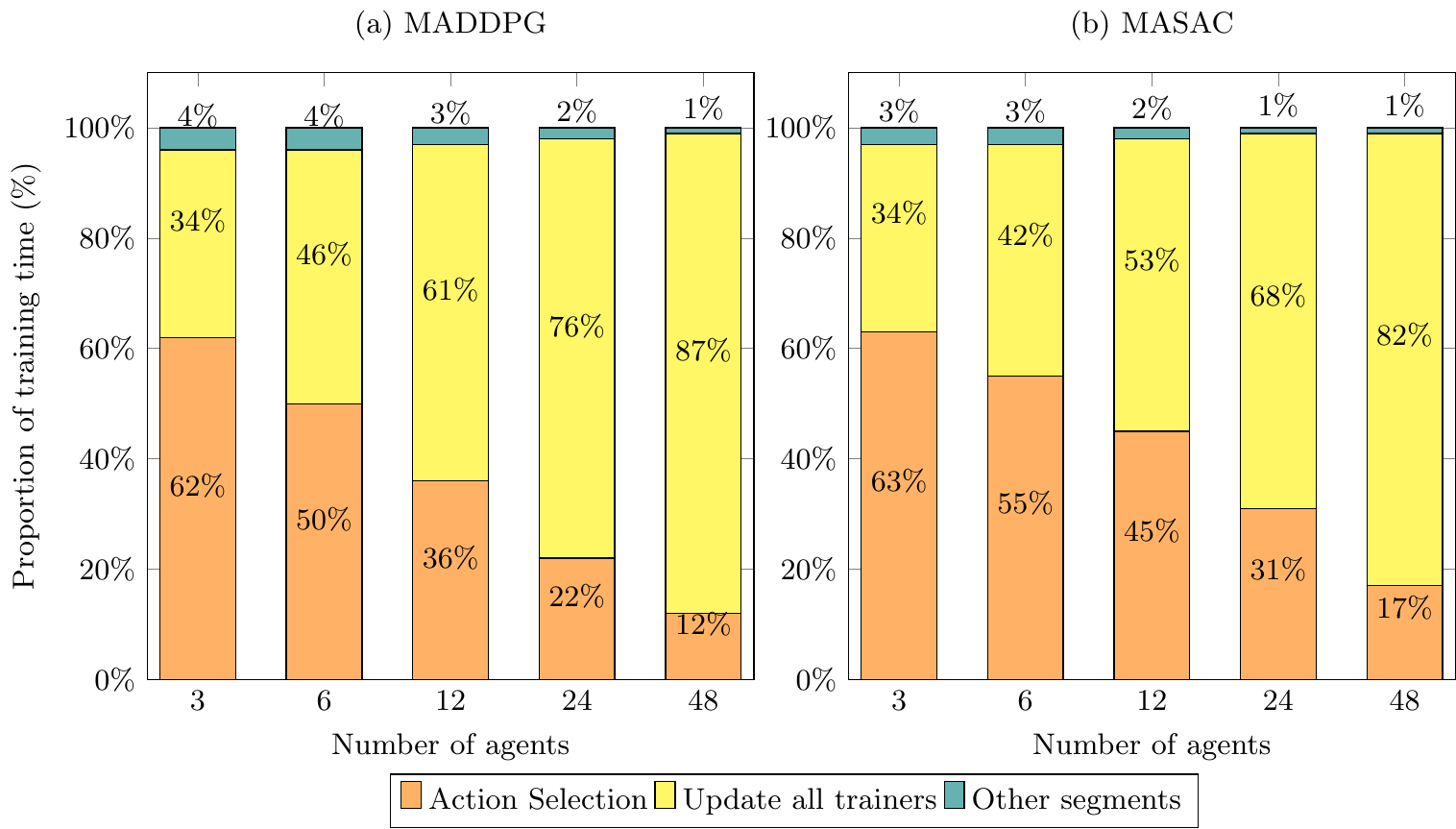}
    \caption{Training time breakdown on Ampere Architecture~RTX 3090 for the MARL workloads~(MADDPG \& MASAC) with 3 to 48 agents. The environment is Predator-Prey.}
  \label{figure2}
\end{figure*}

\begin{figure}[ht]
    \centering
    \pgfplotstableread[col sep=comma]{
X,  Y1, Y2, Y3, Y4
Action Selection, 2.0, 2, 2, 2.1
Update all trainers,  3.3, 3.7, 4.0, 4.3
Total time, 2.8, 3.2, 3.4, 3.9
}\mydata
\begin{tikzpicture}
\begin{axis}[
height=70mm, width=0.5\textwidth,
bar width=0.20,
ybar=0.7pt,
enlarge y limits=0,
enlarge x limits={abs=0.8},
ymin=0,
ymax=8,
y=3.0mm,
legend style={at={(0.5,1.05)}, anchor=south, legend columns=4, font=\footnotesize, cells={anchor=west}},
legend image code/.code={%
            \draw[#1] (0cm,-0.1cm) rectangle (0.5cm,0.2cm);
            },
%ytick distance=5,
%yticklabel style={font=\tiny},
x tick label style = {font = \scriptsize, text width = 2.4cm, align = center, rotate = 360, anchor = north},
ylabel style={align=center, font=\footnotesize}, ylabel={Computation time\\growth rate ($N\times$)},
xtick=data,
xticklabels from table={\mydata}{X},
%y=5mm,
nodes near coords,
nodes near coords style={font=\tiny,
/pgf/number format/.cd,
precision=1,
zerofill,
},
legend style={legend pos=north west,
cells={anchor=west},
font=\tiny,
}
]
\addplot [ybar, fill=orange!60] table [x expr=\coordindex,y=Y1]{\mydata};
\addplot [ybar, fill=yellow!60] table [x expr=\coordindex,y=Y2]{\mydata};
\addplot [ybar, fill=teal!60] table [x expr=\coordindex,y=Y3]{\mydata};
\addplot [ybar, fill=brown!60] table [x expr=\coordindex,y=Y4]{\mydata};
\legend{3 to 6 agents, 6 to 12 agents, 12 to 24 agents, 24 to 48 agents}
\end{axis}
\end{tikzpicture}
%\vspace{-\baselineskip}
%\vspace{-\baselineskip}
    \caption{Computation time growth in MARL modules averaged across the two MARL frameworks.}%. The environment is simple spread.}
    \label{figure3}
\end{figure}
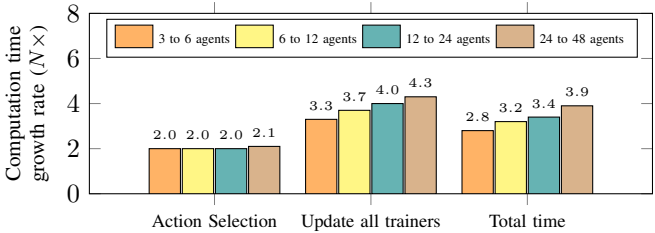

The main contributions of our paper are the following:
\begin{itemize}
     \item We systematically perform a hardware-software performance analysis within the training phases of Multi-agent systems. Further, we present key insights into the performance bottlenecks confronting several key MARL algorithms from a systems perspective.
     \item We explore a neighbor sampling strategy to improve the locality of data access within the \textit{mini-batch sampling} phase. Our preliminary experiments provide performance improvement ranging from $26.66\%$ ($3$ agents) to $27.39\%$ ($12$ agents) in the sampling phase training run-time. Additionally, we achieve $10.2\%$ ($12$ agents) end-to-end training time reduction compared to the SOTA multi-agent algorithms.    
\end{itemize}
\section{Motivation}
\label{sec:motivation}
In multi-agent systems, the training phase is performance-intensive as the agents must collaborate and coordinate to maximize a shared return. Many real-world applications, such as robot fleet coordination~\cite{swamy2020scaled} and traffic light control~\cite{bazzan2009opportunities}, are naturally modeled as multi-agent problems, but they become intractable with the growing number of agents due to the expensive computation required to estimate other agents' policies at each state and a huge amount of neural network parameters. This problem limits their adoption to real-world systems due to their computationally expensive nature and allows them only to deal with a few agents~\cite{zhou2019factorized, liu2020multi}. Figure~\ref{figure2} shows the run-time breakdown of the training phase. We omit the \textit{agents interactions} phase since it primarily depends on environment complexity. 
\textit{Update all trainers} contributes to $\approx$35\% to $\approx$85\% of the training time as the number of MARL agents grows from 3 to 48. This is mainly due to two reasons:~\circled{1}~In MARL, each agent has its own actor and critic networks since they may have different rewards. Each agent must randomly collect a batch of transitions from all other agents to update the critic and actor networks.~\circled{2} The dynamic memory requirements of observation and action spaces also grow quadratically due to each agent having to coordinate with other agents towards sharing their observations and actions. \textit{Action selection} phase occupies a small portion and scales linearly with the number of agents~(Figure~\ref{figure3}). This is because, in \textit{Action selection}, agents consider individual policies to obtain local actions.~\textit{Other segments} include experience collection, reward collection, and policy initialization, and they add a negligible overhead.

\section{Background}
Typically, MARL settings with $N$ agents is defined by a set of states, $S = S_{1} \times ... \times S_{N}$, a set of actions $A = A_{1} \times ... \times A_{N}$. %and a set of observation functions $O = [O_{1}, O_{2}, ..., O_{N}]$ for each agent $i$. 
Each agent selects its action by using a policy $\pi_{\theta_{i}} : O_{i} \times A_{i} \rightarrow [0, 1]$. The state transition~($T : S \times A_{1} \times A_{2} \times ... \times A_{N}$) function produces the next state $S^{'}$, given the current state and actions for each agent. The reward, $R_{i} : S \times A_{i} \rightarrow \mathbb{R}$ for each agent is a function of global state and action of \textit{all other agents}, with the aim of maximizing its own expected return $R_{i} = \sum_{t=0}^{T} \gamma^{t}r_{i}^{t}$, where $\gamma$ denotes the discount factor and $T$ is the time horizon. For this, we use the actor-critic methods such as MADDPG~\cite{lowe2017multi}, %MATD3~\cite{ackermann2019reducing}, 
 MASAC~\cite{haarnoja2018soft}. 

MADDPG~\cite{lowe2017multi} is centralized training and decentralized execution~(CTDE) algorithm mainly designed for mixed environments. Each agent learns an individual policy that maps the observation to its action to maximize the expected return, which is approximated by the critic. MADDPG lets the critic of agent $i$ to be trained  by minimizing the loss with the \textit{target Q-value} and $y_i$ using $\mathcal{L}(\theta_{i}) = {\rm I\!E}_{D}[(Q_{i}(S,A_{1},...A_{n}) - y_{i}^{2}]$, and  $y_{i}=r_{i} + \gamma \overline Q_{i}(S^{'},A_{1}^{'},...A_{n}^{'})_{a_{j}^{'}=\overline \pi(o_{j}^{'})}$, where $S$ and $A_{1},...A_{n}$ represent the joint observations and actions respectively. $D$ is the experience replay buffer that stores the \textit{observations, actions, rewards, and new observations} samples of all agents obtained after the training episodes. The critic networks are augmented with states and actions of all agents to reduce the variance of policy gradients and improve performance. The MARL framework has four networks- actor, critic, target actor, and target critic. $\overline Q_{i}$ and $\overline \pi(o_{j}^{'})$ are the target networks for the stable learning of critic~($Q_{i}$) and actor networks. The target actor estimates the next action from the policy using the state output by the actor network. The target critic aggregates the output from the target actor to compute the target Q-values, which helps to update the critic network and assess the quality of actions taken by agents. The target networks are created to achieve training stability. Note that the updating sequence of networks in the back-propagation phase is critics, actors, then the target networks. 
 
% \textit{MATD3}~\cite{ackermann2019reducing} uses the twin delayed double-centralized critic model, reducing the over-estimation bias problem~\cite{ackermann2019reducing}. The only difference between MATD3 and MADDPG is the use of double-centralized critics and the addition of small amounts of noise to the actions sampled from the buffer.
 
Similar to MADDPG, the centralized critic is introduced in Soft Actor-Critic~(SAC~\cite{haarnoja2018soft}) algorithm. MASAC uses the maximum entropy RL, in which the agents are encouraged to maximize the exploration within the policy. MASAC assigns equal probability to nearly-optimal actions which have similar state-action values and avoids repeatedly selecting the same action. This learning trick will increase the stability, policy exploration, and sample efficiency~\cite{iqbal2019actor, haarnoja2018soft}.

\section{Evaluation Setup}\label{sec:Evaluation_setup}
\textbf{Benchmark.}~Table~\ref{table-1} provides the behavior of selected Multi-agent Particle Environments~(MPE~\cite{lowe2017multi}). We profile and characterize two state-of-the-art MARL algorithms, MADDPG and MASAC. A two-layer ReLU MLP parameterizes the actor and critic networks with 64 units per layer, and the mini-batch size is 1024 for sampling the transitions. In our experiments, we use Adam optimizer~\cite{kingma2014adam} with a learning rate of 0.01, maximum episode length as 25 (max episodes to reach the terminal state), and $\tau$ = 0.01 for updating the target networks. $\gamma$ is the discount factor which is set to 0.95. The size of the replay buffer is $10^{6}$, and the entropy coefficient for MASAC is 0.05. The network parameters are updated after every 100 samples are added to the replay buffer.

\begin{table}[!h]
\caption{\label{table-1}Multi-agent Particle environment.}
%\vspace{-\baselineskip}
\begin{tabular}{
  %|p{\dimexpr.10\linewidth-2\tabcolsep-1.3333\arrayrulewidth}% column 1
  |p{\dimexpr.25\linewidth-2\tabcolsep-1.3333\arrayrulewidth}% column 2
  |p{\dimexpr.75\linewidth-2\tabcolsep-1.3333\arrayrulewidth}|% column 3
  }
\hline
\textbf{Environment} & \textbf{Details}\\
\hline
Cooperative navigation & \textit{N} agents move in a cooperated manner to reach \textit{L} landmarks and the rewards encourages the agents get closer to the landmarks.\\
\hline
Predator-Prey & \textit{N} predators work cooperatively to block the way of \textit{M} fast paced prey agents. The prey agents are environment controlled and they try to avoid the collision with predators.\\
% \hline
%  a number & a number\\
\hline
\end{tabular}
\end{table}

\textbf{Profiling Platform.}~MARL algorithms are implemented with state-of-the-art CPU-GPU compatible TensorFlow-GPU~(v2.11.0). The server runs on Ubuntu Linux~20.04.5~LTS operating system with CUDA~9.0, cuDNN~7.6.5, PCIe Express®~v4.0 with NCCL~v2.8.4 communication library. The machine supports python 3.7.15, TensorFlow-Slim~(v1.1.0) and OpenAI GYM~(v0.10.5).~All the workloads are profiled on single Nvidia GeForce~RTX~3090 Ampere Architecture with Perf~\cite{perf} and NVProf to profile hardware performance counters for performance analysis. Finally, we trained for 60K episodes using default hyper-parameters recommended by the algorithms.

\section{Preliminary Evaluation}\label{sec:Observation and analysis}
This section is organized as follows. First, we present an overview of our profiling result. Then, we divide the computationally dominant functions in \textit{Update all trainers} into multiple modules: \textit{Mini-batch sampling, Target Q calculation}, and \textit{Q loss \& P loss} and present our results in the competitive setting (predator-prey) to understand the key factors limiting MARL in large-scale systems.

\begin{figure*}
    [ht]\centering
    \includegraphics[width=14cm, height=7cm]{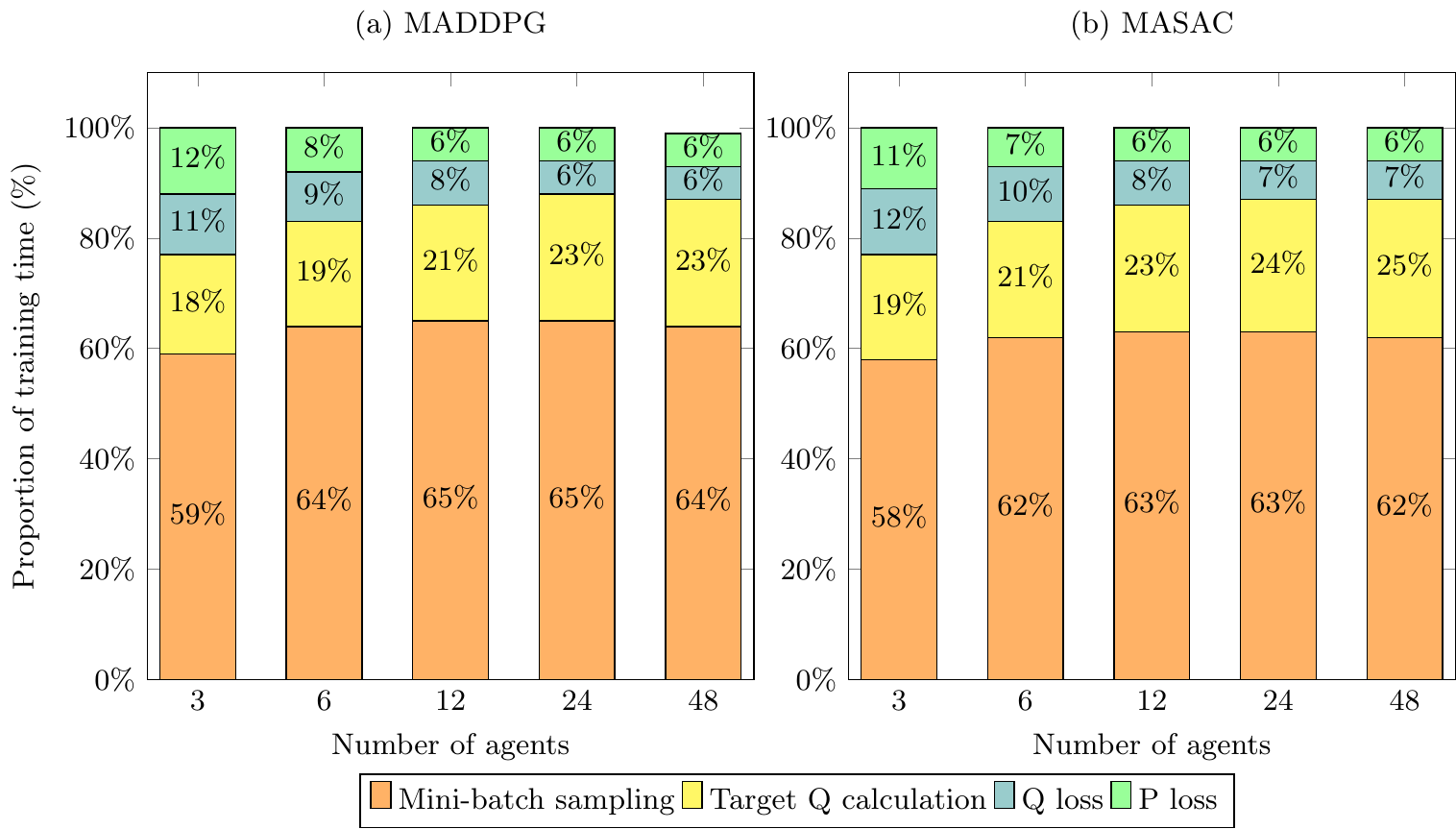}
    \caption{Training time breakdown on Ampere Architecture~RTX 3090 within \textit{Update all trainers} on two different MARL workloads~(MADDPG \& MASAC) with 3 to 48 agents under Predator-Prey environment.}
  \label{figure4}
\end{figure*}

\textbf{Overview of Profile.}\textcolor{black}{~Figure~\ref{figure4} shows the breakdown between the modules, \textit{Mini-batch sampling, Target Q calculation, Q loss, and P loss} that contribute 61\%, 21\%, 10\%, and 8\% to computation time averaging across different workloads, respectively.}
\subsection{Mini-batch sampling} 

Our experimental results in Figure~\ref{figure4} show that mini-batch sampling is the largest time-consuming phase within the {\it Update All Trainers} module. The behavior is also consistent with scaling in other critical hardware performance metrics: \textit{dTLB load misses}-$3.9\times$~(growth rate from $3-6$ agents) and \textit{cache misses}-$3.9\times$~(growth rate from $3-6$ agents).

\textit{Mini-batch sampling} phase is dominated by the collection of random samples from all other agents' replay buffers and updates the parameters of its actor and critic networks. Note that the agent replay buffers are kept separate from each other to capture their past transitions. For each time-step, agent $i$ draws a random index set \{${L_{1}, L_{2},...., L_{K}}$\}~($K$ is the mini-batch size), and first selects $L_1$ to perform a memory lookup in the experience replay buffer to retrieve the corresponding transition and store it in the individual agent buffer. This operation grows as a function of the number of agents, $N$, which is repeated on all $N$ agents. The sampling stage exhibits random memory access patterns and cannot exploit the cache reuse due to randomness in the indices for each agent between the iterations, resulting in increased cache misses. Also, the computation is a heavy burden due to the data accessing from dispersed memory locations, and therefore, the algorithm with large buffer sizes and batch sizes has more cache misses. In cooperative navigation~(\textit{simple spread}~\cite{lowe2017multi}), we observe similar bottlenecks since all the agents are trained together to reach the landmarks while avoiding collisions with each other.

\subsection{Target Q calculation}
The \textit{Target Q calculation} phase is second largest time-consuming phase within {\it Update All Trainers}~(Figure~\ref{figure4}). In this function, each agent performs the \textit{next action calculation, target Q next, and target Q values} as a function of all other agents' joint observation-action space. To calculate the \textit{next action}, the agent $i$ uses its policy network to determine \textit{next action-a'} from the \textit{next state-S'}. \textcolor{black}{In this phase, each agent's policy network involves multiplications with input-weight matrix and additions resulting in performance impact. The obtained a' and S' data are aggregated and concatenated into a single vector in order to compute the \textit{target Q next} amongst the cooperating agents. The input space~(dimension) for the \textit{Q-function} increases quadratically with the number of agents~\cite{sheikh2020multi}.}
The target critic values for each agent $i$ is computed using \textit{target Q next} values from the target actor network. We note that, each agent has to read other agents' policy values; as such for $N$ agents, there is $N\times(N-1)$ memory lookup operations corresponding to the \textit{next action-a'}. 

\subsection{Back-propagation - Q loss \& P loss}  
Back propagation stage is dominated by execution of two networks:~\circled{1}~\textit{critic network} computes the Mean-Squared Error loss between the target critic and critic networks, and \circled{2}~the \textit{actor network} is updated by minimizing the Q values~(computed by the critic network). The total training time increases as the number of agents increases, as shown in Figure~\ref{figure4}. This is because as the number of agents increases, the trainable parameters increase, and \textit{N} policy and \textit{N} critic networks are built for all \textit{N} agents, which incurs extra time to update the weights for each agent. For each update, we sample the random mini-batch of size~(1024 in our studies) transitions from the replay buffer within each agent $i$ from all the other agents and then perform gradient descent on the critic and actor networks.

\section{Neighbor Sampling Strategy}\label{optimization}
From our analysis so far, it can be concluded that the \textit{mini-batch sampling} phase dominates \textit{Update all trainers} phase when the number of agents scales linearly due to each agent sampling all other agents' transition data. Moreover, fetching the transition data from the outlying memory locations significantly affects the overall training time as the problem complexity grows. Among all the hardware metrics, cache misses suffer from the worst scaling factor (3.9$\times$ or higher). Therefore, with the support of loop-level optimization, we explore optimizations that can improve the locality and overall MARL performance. In this paper, we propose an approach at the loop level to optimize the data accessing in the \textit{mini-batch sampling} phase.
\begin{algorithm}
\caption{Neighbor Sampling Strategy}\label{alg:cap}
\begin{algorithmic}[1]
\Require List of random locations $indices$, Replay buffer $\mathcal{D}$ with a size $d$, Neighbors $n$, Batch size $b$
\Ensure Transitions from neighbor $indices$ \newline
% \State obses\_t, actions, rewards, obses\_tp1, dones = [], [], [], [], []
\Comment{\textcolor{teal}{Initialize empty lists, \textit{obses\_t, actions, rewards, obses\_tp1, dones}, to hold observations at time $t$}}
\For{$i$ in $indices$}
%\If{$i$ in {\fontfamily{qcr}\selectfont
%enumerate}($\mathcal{D}$)}
    \State $\alpha \gets$ ${[j | max(0, i - n) ≤ j < min(d, i + n + 1), j \neq i]}$
    \Comment{\textcolor{teal}{$\alpha$ includes all indices in the range (index - neighbors) to (index + neighbors), excluding the current index, and also taking care not to go below 0 or exceed the length of Replay buffer $\mathcal{D}$}}
    %Sample all transitions in $range$ [$i - n$, $i + n$]
    \If{$\alpha$ $\subset$ $\mathcal{D}$} 
        \For{$k$ in $\alpha$}
            \State \textit{obs, act, next\_obs, rew, dones} $\gets$ unpack($\mathcal{D}[k]$) 
            \Comment{\textcolor{teal}{Append these unpacked transition lists to the corresponding existing lists}}
            % \State $\delta$ $\gets$ $\mathcal{D}[j]$
        \EndFor
        % \Comment{Extract all the transitions from the replay buffer $\mathcal{D}$ using the subset of indices $\alpha$}
%         \State \textit{obs, act, next\_obs, rew, dones} $\gets$ {\fontfamily{qcr}\selectfont
% unpack}($\mathcal{D}[j]$) \newline
        % \Comment{Append these transition lists to the corresponding existing lists}
        %\State \Comment{Extract transition data from $\mathcal{D}$}
    \EndIf
%\EndIf
\If{$len(obses\_t)$ $\geq$ $b$}
    \State $break$ \\
\textbf{return} \textit{obses\_t, actions, rewards, obses\_tp1, dones} \newline
\Comment{\textcolor{teal}{Finally, return the corresponding lists converted into NumPy arrays}}
\EndIf
\EndFor
\end{algorithmic}
\label{Algorithm1}
\end{algorithm}

The idea of this approach is to eliminate the computation issues arising due to fetching the data from far away memory locations based on random indices. We investigate the neighbor sampling optimization in MADDPG, where we collectively capture the neighbor transitions of an index $i$ to enable faster data access on a given hardware. Intuitively, at each index $i$, we group the neighbor indices into a single micro-batch and extract the data in a locality-aware memory access order to efficiently sample the transitions. \\

\textbf{Neighbor Sampling Strategy.} Algorithm~\ref{Algorithm1} shows how the \textit{mini-batch sampling} phase selects the neighboring transitions for a random index $i$. We initialize replay buffer $ \mathcal{D}$, neighbors $n$, and batch size $b$. The first loop in line~1 iterates over the random indices. We note that the original computation loop for random sampling is divided into two loops for extracting micro-batched data. The first index $i$ is accessed and checked if $i$ is in the limits of $ \mathcal{D}$ and, if so, we capture the buffer indices from $i - n$ to $i + n$ based on the number of neighbors $n$ and returns a list of neighbors $\alpha$~(line~2) for index $i$. In line~5, output vectors are unpacked and stored as individual vectors in the experience replay tuple consisting of \textit{observations, actions, next observations, rewards, dones}. And these individual vectors are appended to their corresponding parents lists. Finally, all the parent lists which contain the transition data at time-step $t$ are converted as NumPy vectors. Statement 8 checks whether the size of observations has become full~(equal to the batch size $b$); if so, line 10 returns the batch of NumPy vectors.

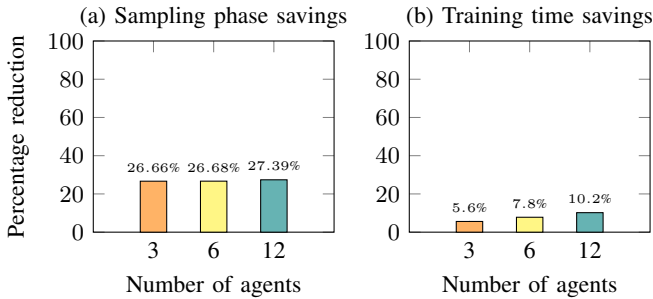
\begin{figure}[ht]
    \raggedright
\begin{tikzpicture}
    \begin{groupplot}[
group style={
    group size=2 by 2,
    ylabels at=edge left,
    horizontal sep=10mm,
            },
    width=0.54\linewidth,
    enlarge x limits=0.5,
    title style = {yshift=-1ex, font=\scriptsize, align=center},
 xlabel={\small Number of agents},
                     ylabel={\small Percentage reduction},
    label style = {font=\small},
     symbolic x coords={3, 6, 12},
    xtick={3, 6, 12},
    ticklabel style={font=\small},
    ymin=0,    ymax=100,
    nodes near coords={\pgfmathprintnumber\pgfplotspointmeta\%},
    every node near coord/.append style={font=\tiny},
    nodes near coords align={vertical},
                    ]
\nextgroupplot[title={\small (a) Sampling phase savings}]
                    \addplot[ybar, fill=orange!60] coordinates {  (3, 26.66)};
                    \addplot[ybar, fill=yellow!60] coordinates { (6, 26.68)};
                    \addplot[ybar, fill=teal!60] coordinates {  (12, 27.39)};
                    %\addplot[ybar, fill=gray] coordinates {  (4, 11.14)};
                    % \addplot[ybar, pattern=north east lines] coordinates {  (5, 15.001)};

                \nextgroupplot[title={\small (b) Training time savings}]
                 \addplot[ybar,fill=orange!60] coordinates {  (3,5.6 )};
                    \addplot[ybar, fill=yellow!60] coordinates { (6, 7.8)};
                    \addplot[ybar, fill=teal!60] coordinates {  (12, 10.2)};
                   % \addplot[ybar, fill=teal] coordinates {  (12-24, 0)};

\end{groupplot}
        \end{tikzpicture}
\caption{(a) Percentage reduction in training time for the \textit{mini-batch sampling} phase for 3, 6 and 12 agents~(MADDPG). (b) Percentage reduction in the \textit{total training time} when the number of agents are scaled by 2$\times$ for MADDPG.~The environment test-bed is Predator-Prey and \textit{Neighbors=3}}
\label{figure6}
\end{figure}
Overall, our optimization improves the cache locality and achieves the performance improvement ranging from $26.66\%$ ($3$ agents) to $27.39\%$ ($12$ agents) during the computationally intensive mini-batch sampling~(Figure~\ref{figure6}).~While studying this optimization, we ensure that there isn't any significant degradation in the mean episode reward.

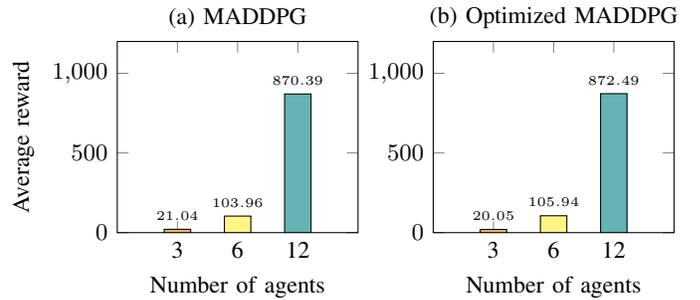
\begin{figure}
    \raggedright
\begin{tikzpicture}
    \begin{groupplot}[
group style={
    group size=2 by 2,
    ylabels at=edge left,
    horizontal sep=10mm,
            },
    width=0.54\linewidth,
    enlarge x limits=0.5,
    title style = {yshift=-1ex, font=\small, align=center},
    xlabel={Number of agents},
    ylabel={Average reward},
    label style = {font=\small},
     symbolic x coords={3, 6, 12},
    xtick={3, 6, 12},
    ticklabel style={font=\small},
    ymin=0,    ymax=1200,
    nodes near coords,
    every node near coord/.append style={font=\tiny},
    nodes near coords align={vertical},
                    ]
\nextgroupplot[title={\small (a) MADDPG}]
                    \addplot[ybar, fill=orange!60] coordinates {  (3, 21.04)};
                    \addplot[ybar,fill=yellow!60] coordinates { (6, 103.96)};
                    \addplot[ybar,  fill=teal!60] coordinates {  (12, 870.39)};
                    %\addplot[ybar, fill=gray] coordinates {  (4, 11.14)};
                    % \addplot[ybar, pattern=north east lines] coordinates {  (5, 15.001)};

                \nextgroupplot[title={\small (b) Optimized MADDPG}]
                 \addplot[ybar, fill=orange!60] coordinates {  (3, 20.05)};
                    \addplot[ybar, fill=yellow!60] coordinates { (6, 105.94)};
                    \addplot[ybar,  fill=teal!60] coordinates {  (12, 872.49)};
                   % \addplot[ybar, fill=teal] coordinates {  (12-24, 0)};

\end{groupplot}
        \end{tikzpicture}
\caption{(a) Average of mean episode rewards of all the agents trained for 60,000 episodes for MADDPG. (b) Average of mean episode rewards all the agents trained for 60,000 episodes after the neighbor sampling optimization for MADDPG. The environment test-bed is Predator-Prey and \textit{Neighbors=3}.} %2$\times$.}% The environment is Cooperative Navigation (simple spread).}
\label{figure7}
\end{figure}

\section{Discussion and Related Work}\label{relatedwork}
Hardware-Software acceleration techniques in RL have been the subject of research in recent years~\cite{babaeizadeh2016ga3c, cho2019fa3c, li2019accelerating, stooke2018accelerated}. For example, to accelerate RL training from the software standpoint, prior works have shown that half-precision~(FP16) quantization can yield significant performance benefits and improve hardware efficiency while achieving adequate convergence~\cite{bjorck2021low}. Other relevant approaches include QuaRL~\cite{krishnan2022quarl}, where quantization is applied to speed up the RL training and inference phases. The authors experimentally demonstrated that quantizing the policies to ≤ 8 bits led to substantial speedups in the training time compared to full precision training. All the prior works differ from our work as they apply quantization to single-agent RL algorithms or neural networks. Further, we explore the neighbor sampling optimization to improve the efficiency of \textit{mini-batch sampling} phase.

Prior studies, like FA3C, have focused on hardware acceleration in multiple parallel worker scenarios, where each agent is controlled independently within their environments using single-agent RL algorithms~\cite{cho2019fa3c}. In contrast, we seek to systematically understand the performance-limiting functions in multi-agent systems, where the agents collaborate in a single shared environment. Agents in such MARL settings usually have high visibility of one another~(leading to large space and action spaces). 

In MARL settings where each agent needs to interact with its neighbor agents, especially in complex environments with lots of observations and huge action spaces, computational bottlenecks may be alleviated using architectural primitives implementing selective attention~\cite{iqbal2019actor, mahajan2021tesseract, ham20203}. As the number of agents increases, the hardware techniques such as near-memory computing could help to perform \textit{mini-batch sampling} efficiently. For the input to critic networks, multi-level data compression~\cite{jain2018gist,zhang2022fast,venkataramani2021rapid} techniques on a targeted group of agents may be used based on their importance in the environment. Also, the cache misses during \textit{mini-batch sampling} phase indicate competition for the LLC cache, which may be addressed through smart cache allocation strategies. Other modules, such as \textit{next action calculation, environment interactions, and action selection phases}, may also benefit from the custom acceleration of key modules.

\section{Conclusion and Future Work}\label{sec:conclusion}
In this work, we present an end-to-end characterization of several popular Multi-Agent Reinforcement Learning algorithms and, in particular, explore the locality-aware neighbor indexing optimization. We find that the \textit{Update all trainers} dominates the training process of MARL algorithms and scales super linearly with the number of agents. Our experimental analysis presents key insights into the modules that are the driving factors behind computational bottlenecks. We also propose an approach at the loop level to optimize the data accessing in the \textit{mini-batch sampling} phase. The proposal achieves performance improvement from $26.66\%$ ($3$ agents) to $27.39\%$ ($12$ agents) in the \textit{mini-batch sampling} phase. Future research includes investigating various pseudo-random sampling strategies and designing a hardware-friendly architecture to fetch the transitions in large-scale MARL efficiently.

\ifCLASSOPTIONcompsoc
  % The Computer Society usually uses the plural form
  \section*{Acknowledgments}
\else
  % regular IEEE prefers the singular form
  \section*{Acknowledgment}
\fi
This research is based on work supported by the National Science Foundation under grant CCF-2114415.

%%%%%%% -- PAPER CONTENT ENDS -- %%%%%%%%

%%%%%%%%% -- BIB STYLE AND FILE -- %%%%%%%%
\bibliographystyle{IEEEtran}
\bibliography{refs}
%%%%%%%%%%%%%%%%%%%%%%%%%%%%%%%%%%%%

\end{document}